\documentclass[apj,iop]{emulateapj}
\usepackage{apjfonts}
\usepackage{xspace}

\newcommand{\jybmkms}{{\ensuremath{\rm Jy\,beam^{-1}\,km\, s^{-1}}}\xspace}
\newcommand{\jybm}{{\ensuremath{\rm Jy\,beam^{-1}}}\xspace}

\newcommand{\ammo}{{\ensuremath{\rm NH_3}}\xspace}

\newcommand{\tkin}{{\ensuremath{\rm T_{kin}}}\xspace}
\newcommand{\csave}{{\ensuremath{\rm c_{s,ave}}}\xspace}

\newcommand{\cc}{{\ensuremath{\rm cm^{-3}}}\xspace}
\newcommand{\kkm}{{\ensuremath{\rm K\,km\, s^{-1}}}\xspace}

\newcommand{\kms}{{\ensuremath{{\rm km\, s^{-1}}}}\xspace}

\begin{document}
\title{EVLA observations of the Barnard 5 star-forming core: embedded filaments revealed}
\shorttitle{Barnard~5:  Filaments revealed}

\shortauthors{J. E. Pineda et al.}
\author{Jaime E. Pineda\altaffilmark{1,2}, 
Alyssa A. Goodman\altaffilmark{3}, 
H\'ector G. Arce\altaffilmark{4}, 
Paola Caselli\altaffilmark{5},  
Steven Longmore\altaffilmark{1}, 
Stuartt Corder \altaffilmark{6,7}
}
\altaffiltext{1}{ESO, Karl Schwarzschild Str. 2, 85748 Garching bei Munchen, Germany}
\altaffiltext{2}{UK ALMA Regional Centre Node, Jodrell Bank Centre for Astrophysics, School of Physics and Astronomy, University of Manchester, Manchester, M13 9PL, UK}
\altaffiltext{3}{Harvard-Smithsonian Center for Astrophysics, 60 Garden St., Cambridge, MA 02138, USA}
\altaffiltext{4}{Department of Astronomy, Yale University, P.O. Box 208101, New Haven, CT 06520-8101, USA}
\altaffiltext{5}{School of Physics and Astronomy, University of Leeds, Leeds LS2 9JT, UK}
\altaffiltext{6}{North American ALMA Science Center, 520 Edgemont Road, Charlottesville, VA 22903, USA}
\altaffiltext{7}{National Radio Astronomy Observatory, 520 Edgemont Road, Charlottesville, VA 22903, USA}
\email{jaime.pineda@manchester.ac.uk}

\slugcomment{Draft version 2, \today, JEP}

\begin{abstract}
We present a $\sim$6.5\arcmin$\times$8$\arcmin$ Expanded Very Large Array (EVLA) 
mosaic observations 
of the \ammo(1,1) emission in the Barnard~5 region in Perseus, with an angular 
resolution of 6$\arcsec$. 
This map covers the coherent region, where the dense gas presents subsonic 
non-thermal motions (as seen from single dish observations with the Green Bank Telescope, GBT). 
The combined EVLA and GBT observations reveal, for the first time, a striking filamentary 
structure (20$\arcsec$ wide or 5,000~AU at the distance of Perseus) in this low-mass 
star forming region. 
The integrated intensity profile of this structure is consistent with models of an isothermal 
filament in hydrostatic equilibrium. 
The observed separation between the B5--IRS1 young stellar object (YSO), in the central 
region of the core, and the northern starless condensation matches the Jeans length of the dense gas. 
This suggests that the dense gas in the coherent region is fragmenting. 
The region observed displays a narrow velocity dispersion, where most of the gas shows
evidence for subsonic turbulence, and where little spatial variations are present.
It is only close to the YSO where an increase in the velocity
dispersion is found, but still displaying subsonic non-thermal motions.
\end{abstract}


\section{Introduction}
Molecular clouds (MCs) observed using low-density tracers 
display velocity dispersions much larger than the thermal value 
\citep[e.g.,][]{Zuckerman+Evans:1974,Larson_1981-turbulence_MC,Myers_1983-subsonic_turbulence}.
These supersonic motions are usually attributed to ``turbulence," and a variety of recent 
numerical models of turbulence can reproduce qualitatively realistic clouds. 
Measurements of MCs' energy budgets show that turbulence \emph{must} be dissipated in 
order for dense cores (where individual or small groups of stars form) to collapse 
and form stars.

Dense cores have been studied using \ammo(1,1) maps, which traces material with densities 
of a few $10^{4}~{\rm cm^{-3}}$. 
It has been found that dense cores show an 
almost constant level of non--thermal motions, within a certain ``coherence'' radius 
\citep{Goodman_1998-coherence}.
The term ``coherent core'' is used to describe the dense gas where non--thermal motions 
are roughly constant, and typically smaller than the thermal motions, independent of scale 
\citep[see also][]{Caselli:2002-n2h+_maps}.

Recent observations carried out with the 100-m Robert F. Byrd Green Bank Telescope (GBT) 
have allowed us to 
create large scale maps of \ammo(1,1) towards four star-forming regions in the 
Perseus Molecular Cloud: B5, IC348--SW, L1448, and L1451 
\citep{Pineda_2010-transition_coherence,Pineda_2011-GBT_Maps}.
One of the main results of \cite{Pineda_2010-transition_coherence} is the clear 
observation (for the first time) of the sharp transition (at $30\arcsec$ resolution) 
between the coherent section of the B5 dense core 
and the more turbulent dense gas outside it 
\citep[see also][for results on the other regions surveyed]{Pineda_2011-GBT_Maps}. 
However, the angular resolution of these observations did not allow us to study in 
great detail the spatial variations of the velocity dispersion or column density.

In this letter, we present new \ammo(1,1) observations of B5 obtained with the 
Expanded Very Large Array (EVLA) 
through the Open Shared Risk Observing (OSRO) program, 
see \cite{Perley_2011-EVLA_description} for a description of the EVLA project. 
These data are combined with previous observations from the 
GBT to study the dense gas traced by \ammo at high angular and 
spectral resolution. 
Here, we present results from the combined EVLA and GBT observations 
that enable us to analyze the kinematic properties of the dense gas traced by \ammo(1,1) 
and also to study the radial profile of new filamentary structure found within this core.

\section{Data}
Observations were carried out with the EVLA of the 
National Radio Astronomy Observatory on May 16th, 2010 (project 10A-181).  
We used the high-frequency K-band receiver and configured the WIDAR correlator  to 
observe a 1\,MHz window around the \ammo(1,1) rest frequency (23.6944950787\,GHz).
The correlator generated 256 channels across the window, giving a 3.90625~kHz 
channel separation, equivalent to $0.049~\kms$  at the observed frequency.
This configuration covers the main hyperfine component and also 
one of the inner pair of satellite lines  for \ammo(1,1).
At this frequency the primary beam of the array is about $1.9\arcmin$.
The array was in the compact (D) configuration, which covers baselines from 35~meters 
up to 1~km, and included 26 antennas. 
The observations covered the entire region of interest using 27 pointings. 
The observations were carried out under fair weather ($\tau_{\rm 22GHz}=0.0689$ nepers) 
for 7 hours.

The bandpass and absolute flux calibrator was the quasar 3C\,48, while the 
phase and amplitude calibrator, quasar J0336+3218, was observed every 15 minutes. 
The $X$-band reference pointing checks were performed every 60 minutes 
using the quasar J0336+3218. 

The raw-data were reduced using CASA image processing software. 
The signal from each baseline was inspected, and baselines showing spurious data were 
removed prior to imaging. 
Each channel was cleaned separately according to the spatial distribution of the emission.
The clean map was imaged using a $6\arcsec$ circular beam and corrected by the primary beam. 
The images were created using multi-scale clean (with scales of 4 and 12~arcsec and smallscalebias=0.2) 
with a robust parameter of $0.5$.

Imaging of the EVLA data only produced a resulting data cube where strong negative bowls were found. 
These negative bowls are generated by the large scale \ammo(1,1) emission which 
arises from angular scales not sampled by the interferometer. 
We included the \ammo(1,1) single dish data obtained with the GBT 
\citep{Pineda_2010-transition_coherence} 
as a prior (or model) during the imaging process to recover the large scale emission. 
The final noise achieved is 14\,m\jybm per channel. 

The integrated intensity map (see Figure~\ref{fig-w11}) shows a few regions 
with negative emission, which would suggest the presence of imaging artifacts. 
However, careful inspection of the data confirms that the spectra towards those positions do not 
show any clear negative bowl associated with imaging artifacts, 
in fact, these regions are about the $-5$-$\sigma$ level still consistent with the map noise.

\section{Results}

\begin{figure*}[h]
\plotone{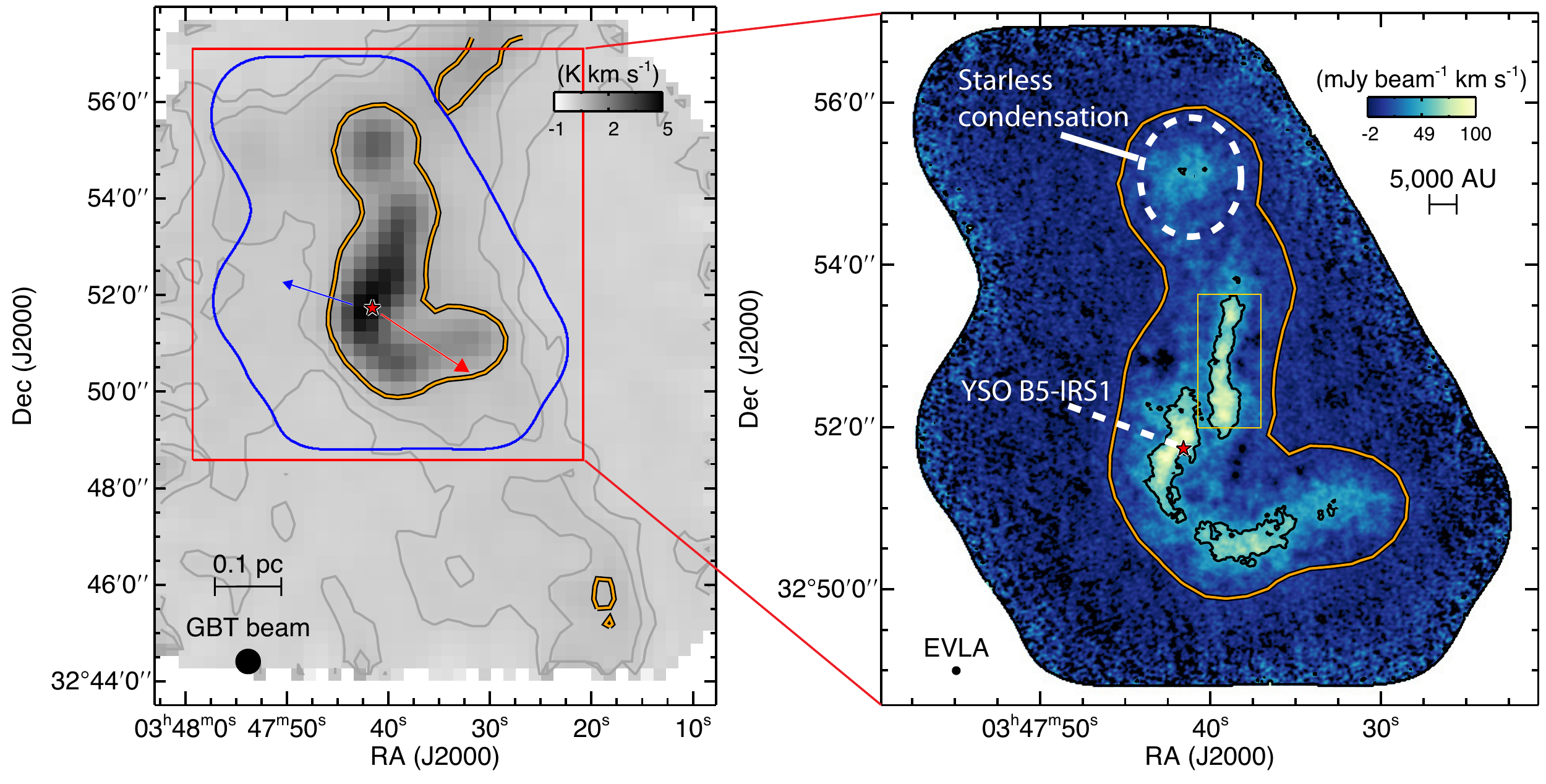}
\caption{\emph{Left panel:}  Integrated intensity map of B5 in \ammo(1,1) obtained with GBT. 
Gray contours show the 0.15 and 0.3~\kkm level in \ammo(1,1) integrated intensity. 
The orange contours show the region in the GBT data where the non-thermal velocity dispersion 
is sub-sonic.
The young star, B5--IRS1, is shown by the star in both panels. 
The outflow direction is shown by the arrows. 
The blue contour shows the area observed with the EVLA, and the 
red box shows the area shown in the right panel.
\emph{Right panel:} Integrated intensity map of B5 in \ammo(1,1) obtained combining 
the EVLA and GBT data. 
Black contour shows the 50\,m\jybmkms level in \ammo(1,1) integrated intensity. 
The yellow box shows the region used in Figure~\ref{fig-fil-prop}. 
The northern starless condensation is shown by the dashed circle.
\label{fig-w11}}
\end{figure*}

Left panel of Figure~\ref{fig-w11} shows the integrated intensity map obtained using the GBT 
at 30$\arcsec$ resolution. 
Gray contours in Figure~\ref{fig-w11} show the extent of \ammo(1,1) emission.
The orange contours show the regions in the GBT data within which the non-thermal velocity dispersion 
is sub-sonic.
The blue contour shows the region observed with the EVLA and presented in the right panel, 
and it covers the entire central region where the sub-sonic non-thermal 
velocity dispersions are observed.
The resulting \ammo(1,1) integrated intensity map for B5 is shown in right panel of 
Figure~\ref{fig-w11}, and it covers 
a region of size 6.5$\arcmin\times$8$\arcmin$.

The new high resolution integrated intensity map (right panel of Figure~\ref{fig-w11}) reveals 
that within the region of subsonic non-thermal motions found in the single dish data (orange contour) 
filamentary structures appear. 
These filaments are narrow, with widths of  $\approx20\arcsec$ or 5,000~AU at the 
distance of Perseus \citep[250~pc;][]{Hirota_2008-NGC1333_Distance}.

\begin{figure*}[h]
\plotone{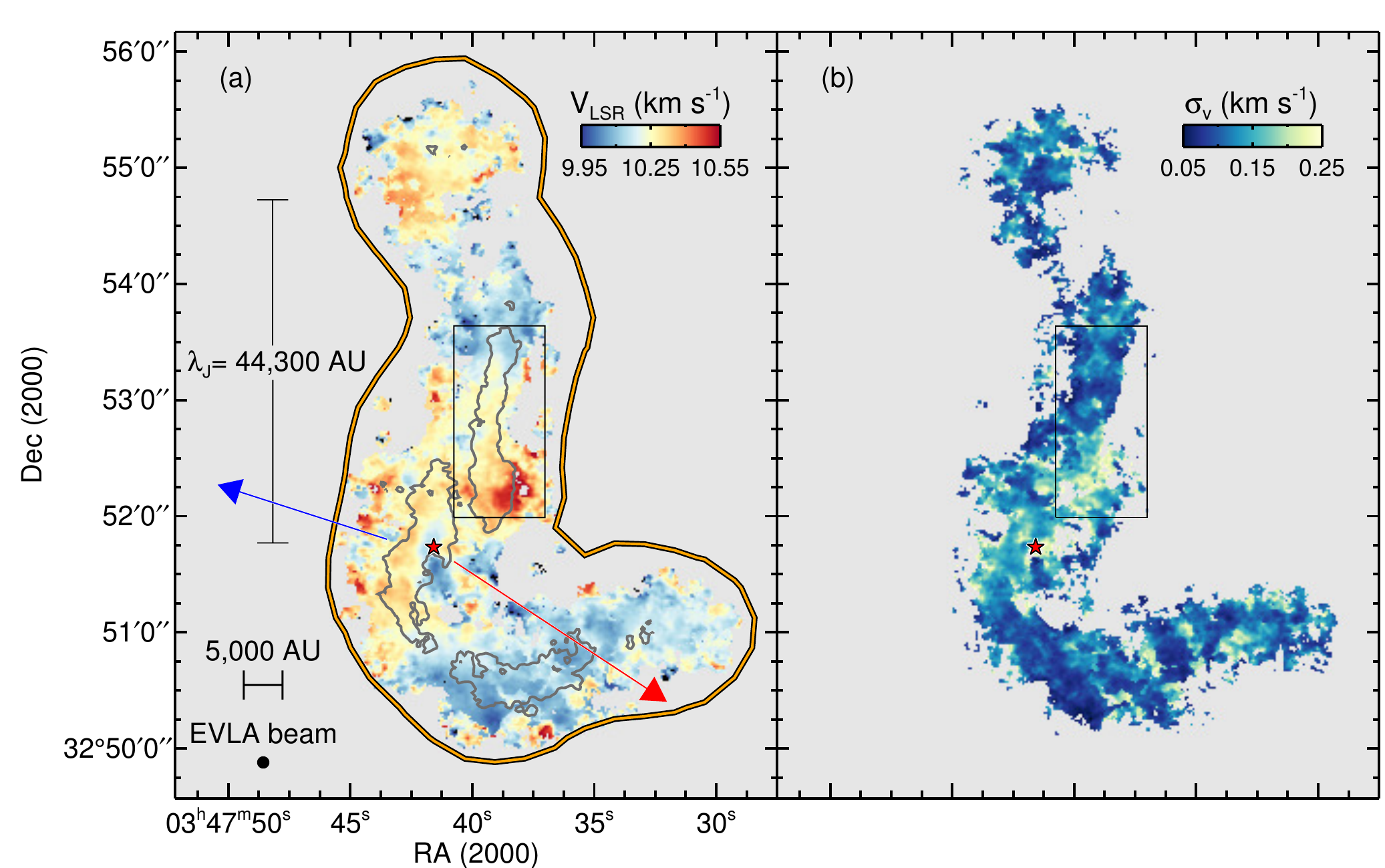}
\caption{ {\bf (a)} 
Centroid velocity map of B5 obtained by fitting all observed \ammo hyperfine components simultaneously.
The position of the YSO is shown by the star, and the orientation of the outflow is shown 
by the arrows. 
The estimated Jeans length, $\lambda_{J}$, of the dense gas is shown by the vertical scale bar, 
see Section~\ref{section-concl}.
The EVLA beam size is shown at the bottom left. 
The orange contour shows the region with subsonic non-thermal motions identified from the GBT data. 
Gray contour shows the 50\,m\jybmkms level in \ammo(1,1) integrated intensity, same as 
right panel of Figure~\ref{fig-w11}. 
The Black box shows the region used in Figure~\ref{fig-fil-prop}. 
{\bf (b)} 
Velocity dispersion map derived from fitting all \ammo hyperfine components simultaneously.
\label{fig-vc-dv}}
\end{figure*}

An important feature of molecular line observations is the ability to probe the kinematics of the gas. 
Here we fit simultaneously all hyperfine components of the \ammo(1,1) line using a forward model 
previously presented by \cite{GBT:Perseus}, see also \cite{Pineda_2010-transition_coherence}. 
This method describes the emission at every position with a  
centroid velocity ($v_{LSR}$), velocity dispersion ($\sigma_{v}$), 
kinetic temperature ($T_{k}$), excitation temperature ($T_{ex}$) and opacity ($\tau_{11}$), 
while also including the response of the frequency channel using a sinc profile. 
Since the kinetic temperature is only used to predict the \ammo(2,2) line 
(not observed due to the constraints in the OSRO program), it does not have any 
effect on the remaining parameters. 
We use a fixed value of the kinetic temperature of 
10~K for the entire region, which is consistent with the results obtained by 
\cite{Pineda_2011-GBT_Maps} using the single dish data. 
If the resulting fit does not provide an accurate velocity dispersion,
$\sigma_{v} < 0.05\,\kms$ (velocity dispersion narrower than the expected thermal value 
for gas at 6~K) or 
$\sigma_{\sigma_{v}} > 0.2\,\sigma_{v}$ (signal-to-noise for the velocity dispersion less than 5), 
the fit is repeated but with a 
fixed value of 5~K for excitation temperature. 
The centroid velocity and velocity dispersion maps are shown in Figure~\ref{fig-vc-dv}.

The centroid velocity map shows little variation ($<0.6$\,\kms) across the entire region, 
see Figure~\ref{fig-vc-dv}a. 
The region close to the YSO displays a velocity gradient at a position angle that is close 
to the outflow (shown by the arrows) but in opposite direction. 
The velocity dispersion map, see Figure~\ref{fig-vc-dv}b, shows vast regions where 
small and uniform velocity dispersion are found. 
It is only towards positions close to the YSO where slightly broader lines are found.

\begin{figure}[h]
\plotone{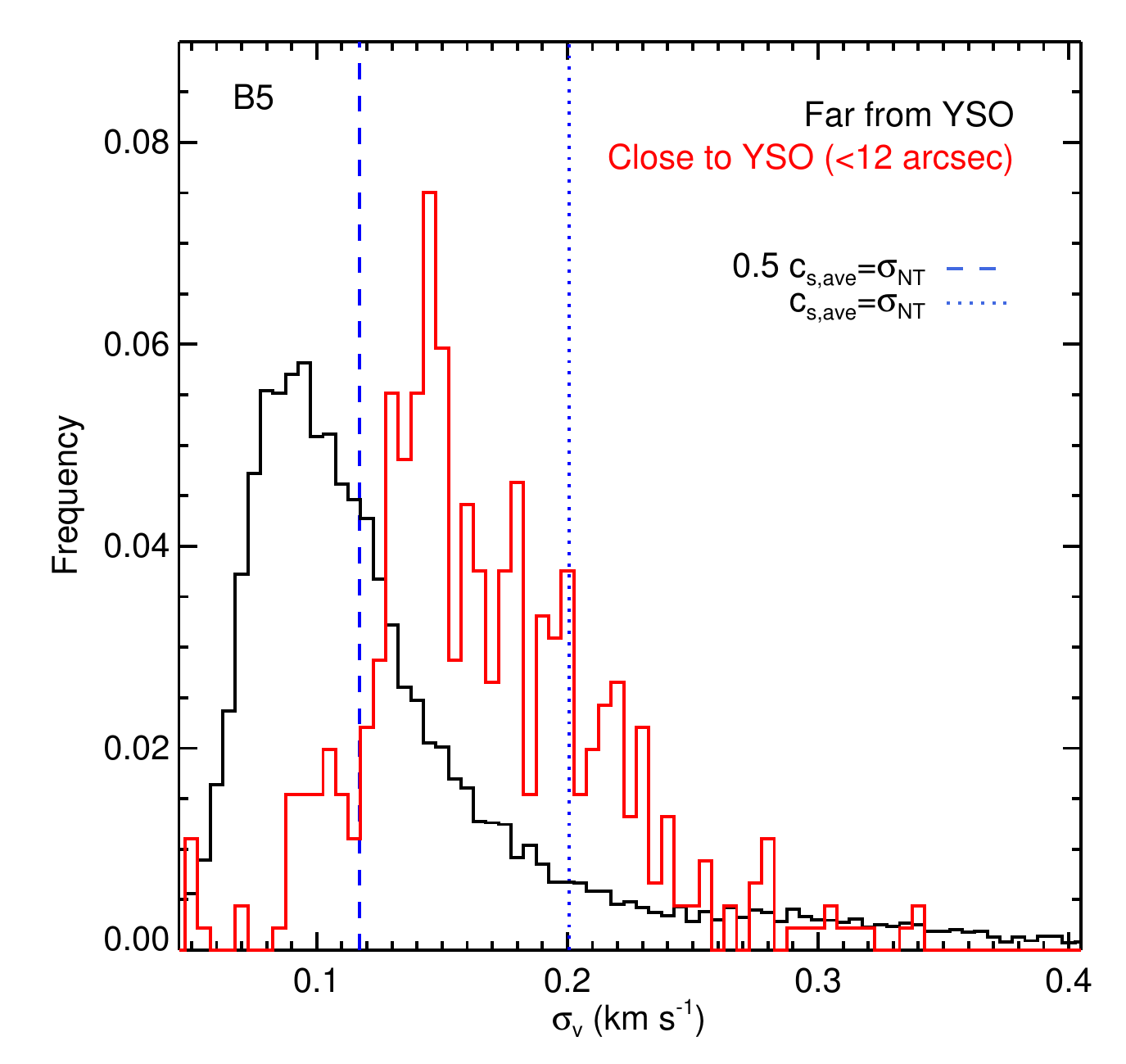}
\caption{Velocity dispersion distribution for positions at a distance to B5--IRS1 larger (smaller) 
than $12\arcsec$ is shown in black (red). 
Blue vertical lines show the expected velocity dispersion for two values of the velocity dispersion 
non-thermal component ($\sigma_{NT}$): 
$0.5\, \csave$, and $\csave$, where $\csave$ is the sound speed of the average particle 
($\mu=2.33$) assuming \tkin=10~K. 
The velocity dispersions, $\sigma_{v}$, obtained far enough from the YSO (black histogram) 
are consistently below the values expected for supersonic non-thermal motions. 
Positions close to the YSO (red histogram) display larger velocity dispersions than the 
rest of the region, but they are still consistent with subsonic non-thermal motions.
\label{fig-hist-dv}
}
\end{figure}

Figure~\ref{fig-hist-dv} presents the distribution of the derived velocity dispersion towards B5. 
Two histograms are shown Figure~\ref{fig-hist-dv} depending on the proximity to the YSO in B5: 
(a) positions close to the YSO (separated by 2 beams or less, $<$12$\arcsec$) in red, and 
(b) all other pixels in black. 
For points far from the YSO the velocity dispersions are small and the velocity dispersion 
distribution (black histogram in Figure~\ref{fig-hist-dv}) is narrow, almost every pixel at a distance larger 
than 2 beams (12\arcsec) from the central YSO displays sub-sonic non-thermal motions. 
The velocity dispersion distribution of pixels far from the YSO, 
black histogram in Figure~\ref{fig-hist-dv}, 
peaks at a value lower than  what is expected if the non-thermal component, $\sigma_{NT}$, 
is equal to half the thermal velocity dispersion,  $0.5\, \csave$.
The velocity dispersion of points close to the YSO, red histogram in Figure~\ref{fig-hist-dv}, 
presents lines broader than the rest of the core, but they still display velocity dispersions 
with a subsonic non-thermal component. 
This increase in the velocity dispersion might be the effect of the radiation from the embedded 
YSO 
or  due to the interaction between the outflow or stellar wind and the dense gas.

Herschel observations of the IC~5146 star forming region 
\citep{Arzoumanian_2011-Herschel_Filaments_IC5146} revealed filamentary structure 
seen in the column density maps 
\citep[see also][]{Andre_2010-Herschel_GBS_Filaments}.
\cite{Arzoumanian_2011-Herschel_Filaments_IC5146} fitted the column density profile 
of filaments with a cylindrical filament model, 
\begin{equation}
\Sigma(r) = A_{p}\frac{\rho_{c}\,R_{flat}}{\left(1+(r/R_{flat})^2\right)^{(p-1)/2}}~,
\label{eq-col-dens}
\end{equation}
where $\rho_{c}$ is the filament's central density, 
$r$ is the cylindrical radius, 
$p$ is the power-law density exponent at large radii, 
$R_{flat}$ is the radius of the density profile inner flat section, and 
$A_{p}$ is a finite constant factor dependent on $p$ and the filament inclination angle. 
They find filaments which are well fit with a density exponent in the range $p=1.5-2.5$  
\citep[see also][]{Lada_1999-IC5146_Structure}.
These exponents do not agree with the predicted values of  an isothermal filament 
in hydrostatic equilibrium where a steeper exponent, $p=4$, is expected 
\citep{Ostriker_1964-Filament_Model}.

\begin{figure}[h]
%\epsscale{0.7}
\plotone{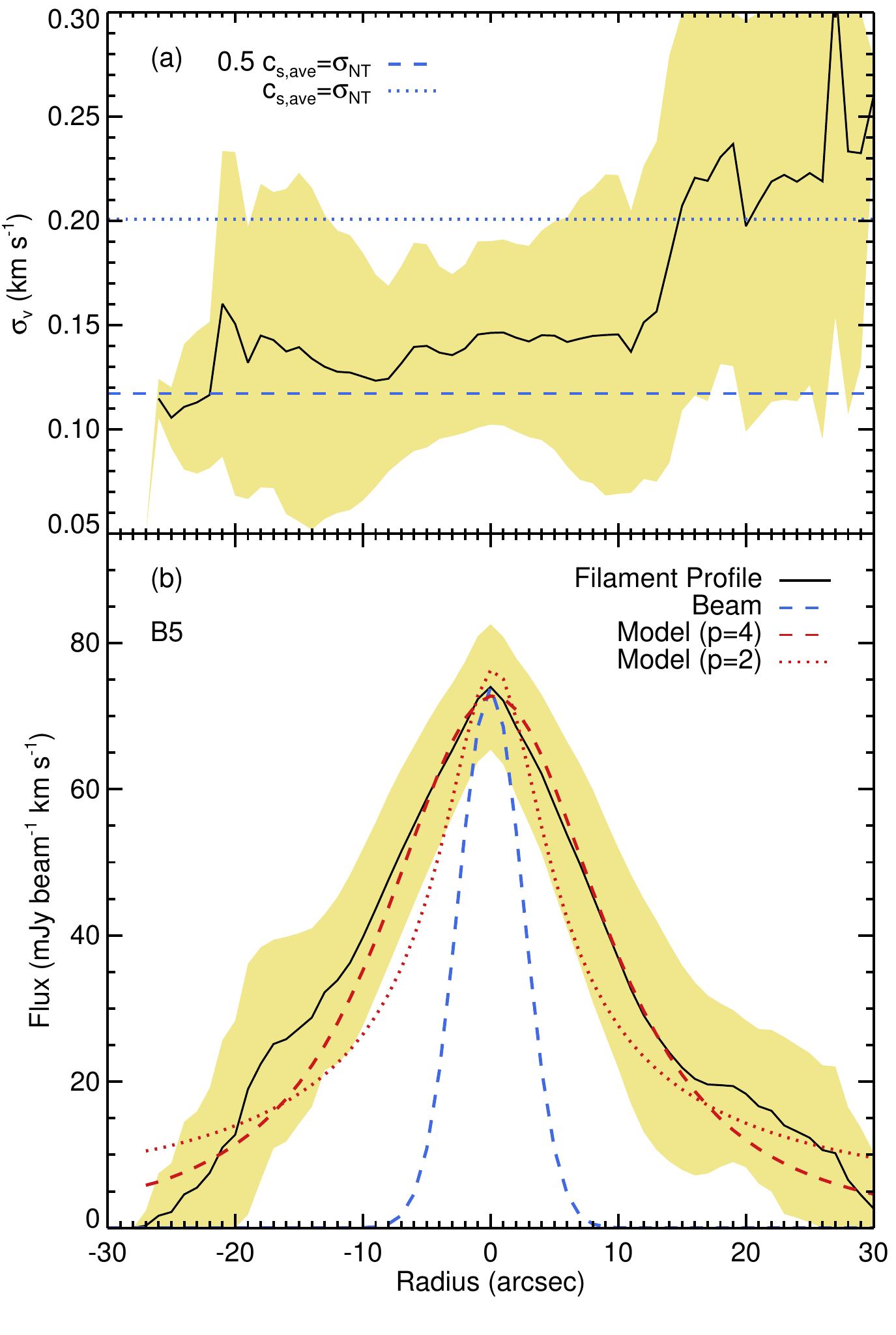}
\caption{ {\bf (a)} 
Average velocity dispersion profile perpendicular to the filament shown by the 
yellow box in Figure~\ref{fig-w11}. 
The dispersion of the radial profile along the filament is shown by the yellow area.
Similar to Figure~\ref{fig-hist-dv}, blue horizontal lines show the expected velocity dispersion for 
two values $\sigma_{NT}$:  $0.5\, \csave$, and $\csave$. 
{\bf (b)} 
Average integrated emission profile perpendicular to the filament, with the dispersion shown in yellow. 
The beam response is shown by the blue line, while two filament models are shown by the red curves. 
\label{fig-fil-prop}}
\epsscale{1}
\end{figure}

Here we focus our attention towards the filament shown in Figure~\ref{fig-w11} by the yellow box. 
Since this filament is almost perfectly aligned in the North-South direction we perform a series 
of horizontal cuts, and define the radius as the distance from the peak at a given cut. 
The average velocity dispersion and integrated intensity emission profiles along the filament are 
shown in panels (a) and (b) of Figure~\ref{fig-fil-prop}, respectively, where the spread in 
the distribution is shown by the yellow area. 
Figure~\ref{fig-fil-prop}a shows that the velocity dispersion does not change across the filament, 
and it is consistent with subsonic non-thermal velocity dispersions (delimited by the dotted line). 
Figure~\ref{fig-fil-prop}b shows the integrated intensity profile, which is much wider than the 
beam of the combined EVLA and GBT data (shown by the blue dash line). 
Since we do not observe the \ammo(2,2) with the EVLA we cannot provide a column density profile, 
but we use \ammo(1,1) integrated intensity as a proxy for the total column density 
\citep{GBT:Perseus,Friesen_2009-GBT_Oph,Foster_2009-GBT,Pineda_2011-GBT_Maps}. 
The integrated intensity profile is fitted with a model which follows equation~\ref{eq-col-dens}, 
and the best fit results for $p=2$ and $p=4$ are shown in Figure~\ref{fig-fil-prop}. 
The figure shows that this filament in B5, in opposition to what is found by 
\cite{Arzoumanian_2011-Herschel_Filaments_IC5146}  on larger filaments, 
is better fit by an isothermal hydrostatic filament model 
(see also Hacar et al., submitted, or Bourke et al., submitted, for other examples of objects with 
similar column density profiles).

\section{Discussion and conclusion}\label{section-concl}

The observations presented here show that subsonic non-thermal velocity dispersions 
display little variations across the region of coherence, and even across a filament. 
Also, the coherent region is far from presenting an uniform column density (as 
traced by the \ammo(1,1) integrated intensity), as filamentary substructures are revealed. 
But the presence of substructures within the region of coherence should not be surprising, 
the lack of turbulent support might allow external forces or Jeans-like instabilities
to generate over-densities without 
much difficulties.

Filaments have received special attention recently. 
They appear as a natural outcome from turbulent simulations of molecular clouds 
\citep[e.g.,][]{Klessen_2004-Gravoturbulence_Clusters}, and 
also from the fragmentation of a modulated layer \citep{Myers_2009-Filaments}. 
Recent Herschel observations have shown that filaments are 
commonly found in star-forming regions 
\citep{Andre_2010-Herschel_GBS_Filaments,Arzoumanian_2011-Herschel_Filaments_IC5146}. 
However, these filaments present different properties than those presented here : 
(a) they are bigger structures, with a characteristic width of $\sim$0.1~pc; and 
(b) they are not well fitted by an hydrostatic and isothermal cylindrical model. 
The filamentary structures revealed by our observations appear to have different 
properties than filaments previously studied.  
The filaments in B5 are embedded in a dense region with subsonic turbulence, 
and the filament shown in Figure~\ref{fig-fil-prop} is better fitted with the 
hydrostatic isothermal filament model.
The differences between the filaments identified using Herschel and EVLA are due to physical differences 
between the structures. 
The Herschel identified filaments are found at low column densities and have a characteristic size of $\sim0.1$\,pc, 
which is similar to the size of the coherent structures found in single dish observations 
\citep{Pineda_2010-transition_coherence,Goodman_1998-coherence}, 
but much smaller than the largest scales detected ($\sim1\deg$).
While the filaments revealed by the EVLA observations are substructures 
within the coherent region which are also at higher column densities than the structures 
found in Herschel. 
Also, since the filament width (see Figure~\ref{fig-fil-prop}b) is smaller than the EVLA primary beam 
and the single dish data is used to recover the extended emission then the radial profile is little 
affected by the spatial filtering of interferometric observations.

Fragmentation in the coherent region should occur at the Jeans length, 
\begin{equation}
\lambda_{J}=44,300~ \left(\frac{T}{10\,{\rm K}}\right)^{1/2}  
                                        \left(\frac{\langle n\rangle}{10^{4}\,\cc}\right)^{-1/2}
                                        \left(\frac{\mu}{2.3}\right)^{-1}  ~{\rm AU}\,,
\end{equation}
where $T$ is the temperature of the gas, 
$\langle n\rangle$ is the average density, and 
$\mu$ is the mean molecular weight. 
For the average values of density and temperature in B5 from the dense gas 
\citep[$T=10$~K and $\langle n\rangle=10^{4}\,\cc$,][]{Pineda_2011-GBT_Maps} the Jeans length 
correspond to the separation between the B5--IRS1 YSO and the starless condensation 
in the northern part of the region observed. 
Fragmentation has been argued to explain the stars separation in young star-forming regions 
\citep[e.g.,][]{Hartmann_2002-Fragmentation_Taurus,Teixeira_2006-Thermal_Fragmentation,Teixeira_2007-NGC2264_SMA} 
and the mass function of cores or condensations \citep[e.g.,][]{Testi_1998-Serpens_IMF,Motte_1998-Oph_CMF,Andre_2007-Oph_N2H+}. 
Here we show evidence for fragmentation occurring within the coherent region, 
which is not necessarily linked to the filament formation process.

These results support the idea that a coherent region needs to be created to form a low-mass star. 
It is in this coherent region where the dense gas that has lost its turbulent support accumulates, and 
then it can easily fragment, create a filament,  and/or undergo gravitational collapse to form a star.
We expect that future EVLA and ALMA observations of star forming regions will allow us to produce 
temperature and velocity dispersion maps which will allow us to perform a much better job 
on characterising the physical properties of filaments and of these coherent regions.

\acknowledgments
We would like to thank Alvaro Hacar for discussions regarding the properties of filaments. 
The Expanded Very Large Array is operated by the National Radio
Astronomy Observatory. The National Radio Astronomy Observatory
is a facility of the National Science Foundation, operated
under cooperative agreement by Associated Universities, Inc.
This material is based upon work supported by the National Science Foundation under Grant 
AST-0908159 to AAG and AST-0845619 to HGA. 

\facility{{\it Facilities:} EVLA, GBT}


\begin{thebibliography}{24}
\expandafter\ifx\csname natexlab\endcsname\relax\def\natexlab#1{#1}\fi

\bibitem[{{Andr{\'e}} {et~al.}(2007){Andr{\'e}}, {Belloche}, {Motte}, \&
  {Peretto}}]{Andre_2007-Oph_N2H+}
{Andr{\'e}}, P., {Belloche}, A., {Motte}, F., \& {Peretto}, N. 2007, \aap, 472,
  519

\bibitem[{{Andr{\'e}} {et~al.}(2010){Andr{\'e}}, {Men'shchikov}, {Bontemps},
  {K{\"o}nyves}, {Motte}, {Schneider}, {Didelon}, {Minier}, {Saraceno},
  {Ward-Thompson}, {di Francesco}, {White}, {Molinari}, {Testi}, {Abergel},
  {Griffin}, {Henning}, {Royer}, {Mer{\'{\i}}n}, {Vavrek}, {Attard},
  {Arzoumanian}, {Wilson}, {Ade}, {Aussel}, {Baluteau}, {Benedettini},
  {Bernard}, {Blommaert}, {Cambr{\'e}sy}, {Cox}, {di Giorgio}, {Hargrave},
  {Hennemann}, {Huang}, {Kirk}, {Krause}, {Launhardt}, {Leeks}, {Le Pennec},
  {Li}, {Martin}, {Maury}, {Olofsson}, {Omont}, {Peretto}, {Pezzuto}, {Prusti},
  {Roussel}, {Russeil}, {Sauvage}, {Sibthorpe}, {Sicilia-Aguilar}, {Spinoglio},
  {Waelkens}, {Woodcraft}, \& {Zavagno}}]{Andre_2010-Herschel_GBS_Filaments}
{Andr{\'e}}, P., {et~al.} 2010, \aap, 518, L102+

\bibitem[{{Arzoumanian} {et~al.}(2011){Arzoumanian}, {Andr{\'e}}, {Didelon},
  {K{\"o}nyves}, {Schneider}, {Men'shchikov}, {Sousbie}, {Zavagno}, {Bontemps},
  {di Francesco}, {Griffin}, {Hennemann}, {Hill}, {Kirk}, {Martin}, {Minier},
  {Molinari}, {Motte}, {Peretto}, {Pezzuto}, {Spinoglio}, {Ward-Thompson},
  {White}, \& {Wilson}}]{Arzoumanian_2011-Herschel_Filaments_IC5146}
{Arzoumanian}, D., {et~al.} 2011, \aap, 529, L6+

\bibitem[{{Caselli} {et~al.}(2002){Caselli}, {Benson}, {Myers}, \&
  {Tafalla}}]{Caselli:2002-n2h+_maps}
{Caselli}, P., {Benson}, P.~J., {Myers}, P.~C., \& {Tafalla}, M. 2002, \apj,
  572, 238

\bibitem[{{Foster} {et~al.}(2009){Foster}, {Rosolowsky}, {Kauffmann}, {Pineda},
  {Borkin}, {Caselli}, {Myers}, \& {Goodman}}]{Foster_2009-GBT}
{Foster}, J.~B., {Rosolowsky}, E.~W., {Kauffmann}, J., {Pineda}, J.~E.,
  {Borkin}, M.~A., {Caselli}, P., {Myers}, P.~C., \& {Goodman}, A.~A. 2009,
  \apj, 696, 298

\bibitem[{{Friesen} {et~al.}(2009){Friesen}, {Di Francesco}, {Shirley}, \&
  {Myers}}]{Friesen_2009-GBT_Oph}
{Friesen}, R.~K., {Di Francesco}, J., {Shirley}, Y.~L., \& {Myers}, P.~C. 2009,
  \apj, 697, 1457

\bibitem[{{Goodman} {et~al.}(1998){Goodman}, {Barranco}, {Wilner}, \&
  {Heyer}}]{Goodman_1998-coherence}
{Goodman}, A.~A., {Barranco}, J.~A., {Wilner}, D.~J., \& {Heyer}, M.~H. 1998,
  \apj, 504, 223

\bibitem[{{Hartmann}(2002)}]{Hartmann_2002-Fragmentation_Taurus}
{Hartmann}, L. 2002, \apj, 578, 914

\bibitem[{{Hirota} {et~al.}(2008){Hirota}, {Bushimata}, {Choi}, {Honma},
  {Imai}, {Iwadate}, {Jike}, {Kameya}, {Kamohara}, {Kan-Ya}, {Kawaguchi},
  {Kijima}, {Kobayashi}, {Kuji}, {Kurayama}, {Manabe}, {Miyaji}, {Nagayama},
  {Nakagawa}, {Oh}, {Omodaka}, {Oyama}, {Sakai}, {Sasao}, {Sato}, {Shibata},
  {Tamura}, \& {Yamashita}}]{Hirota_2008-NGC1333_Distance}
{Hirota}, T., {et~al.} 2008, \pasj, 60, 37

\bibitem[{{Klessen} {et~al.}(2004){Klessen}, {Ballesteros-Paredes}, {Li}, \&
  {Mac Low}}]{Klessen_2004-Gravoturbulence_Clusters}
{Klessen}, R.~S., {Ballesteros-Paredes}, J., {Li}, Y., \& {Mac Low}, M. 2004,
  in Astronomical Society of the Pacific Conference Series, Vol. 322, The
  Formation and Evolution of Massive Young Star Clusters, ed.
  {H.~J.~G.~L.~M.~Lamers, L.~J.~Smith, \& A.~Nota}, 299--+

\bibitem[{{Lada} {et~al.}(1999){Lada}, {Alves}, \&
  {Lada}}]{Lada_1999-IC5146_Structure}
{Lada}, C.~J., {Alves}, J., \& {Lada}, E.~A. 1999, \apj, 512, 250

\bibitem[{{Larson}(1981)}]{Larson_1981-turbulence_MC}
{Larson}, R.~B. 1981, \mnras, 194, 809

\bibitem[{{Motte} {et~al.}(1998){Motte}, {Andre}, \&
  {Neri}}]{Motte_1998-Oph_CMF}
{Motte}, F., {Andre}, P., \& {Neri}, R. 1998, \aap, 336, 150

\bibitem[{{Myers}(1983)}]{Myers_1983-subsonic_turbulence}
{Myers}, P.~C. 1983, \apj, 270, 105

\bibitem[{{Myers}(2009)}]{Myers_2009-Filaments}
---. 2009, \apj, 700, 1609

\bibitem[{{Ostriker}(1964)}]{Ostriker_1964-Filament_Model}
{Ostriker}, J. 1964, \apj, 140, 1056

\bibitem[{{Perley} {et~al.}(2011){Perley}, {Chandler}, {Butler}, \&
  {Wrobel}}]{Perley_2011-EVLA_description}
{Perley}, R.~A., {Chandler}, C.~J., {Butler}, B.~J., \& {Wrobel}, J.~M. 2011,
  ArXiv e-prints

\bibitem[{{Pineda} {et~al.}(2010){Pineda}, {Goodman}, {Arce}, {Caselli},
  {Foster}, {Myers}, \& {Rosolowsky}}]{Pineda_2010-transition_coherence}
{Pineda}, J.~E., {Goodman}, A.~A., {Arce}, H.~G., {Caselli}, P., {Foster},
  J.~B., {Myers}, P.~C., \& {Rosolowsky}, E.~W. 2010, \apjl, 712, L116

\bibitem[{{Pineda} {et~al.}(2011){Pineda}, {Goodman}, {Arce}, {Caselli},
  {Foster}, {Myers}, \& {Rosolowsky}}]{Pineda_2011-GBT_Maps}
---. 2011, \apj, In prep.

\bibitem[{{Rosolowsky} {et~al.}(2008){Rosolowsky}, {Pineda}, {Foster},
  {Borkin}, {Kauffmann}, {Caselli}, {Myers}, \& {Goodman}}]{GBT:Perseus}
{Rosolowsky}, E.~W., {Pineda}, J.~E., {Foster}, J.~B., {Borkin}, M.~A.,
  {Kauffmann}, J., {Caselli}, P., {Myers}, P.~C., \& {Goodman}, A.~A. 2008,
  \apjs, 175, 509

\bibitem[{{Teixeira} {et~al.}(2007){Teixeira}, {Zapata}, \&
  {Lada}}]{Teixeira_2007-NGC2264_SMA}
{Teixeira}, P.~S., {Zapata}, L.~A., \& {Lada}, C.~J. 2007, \apjl, 667, L179

\bibitem[{{Teixeira} {et~al.}(2006){Teixeira}, {Lada}, {Young}, {Marengo},
  {Muench}, {Muzerolle}, {Siegler}, {Rieke}, {Hartmann}, {Megeath}, \&
  {Fazio}}]{Teixeira_2006-Thermal_Fragmentation}
{Teixeira}, P.~S., {et~al.} 2006, \apjl, 636, L45

\bibitem[{{Testi} \& {Sargent}(1998)}]{Testi_1998-Serpens_IMF}
{Testi}, L., \& {Sargent}, A.~I. 1998, \apjl, 508, L91

\bibitem[{{Zuckerman} \& {Evans}(1974)}]{Zuckerman+Evans:1974}
{Zuckerman}, B., \& {Evans}, II, N.~J. 1974, \apjl, 192, L149

\end{thebibliography}
\end{document}